\documentclass[12pt]{article}
\usepackage{graphicx}

\def\institute{Technische Universit\"at Dortmund, Experimentelle Physik 4, Otto-Hahn-Str. 4a, D-44227 Dortmund, Germany}

\def\Title#1{\begin{center} {\Large #1 } \end{center}}
\def\Author#1{\begin{center}{ \sc #1} \end{center}}
\def\Address#1{\begin{center}{ \it #1} \end{center}}

\newcommand\pubblock{\rightline{\begin{tabular}{l} \\ \end{tabular}}}
\newenvironment{Abstract}{\begin{quotation}  }{\end{quotation}}
\newenvironment{Presented}{\begin{quotation} \begin{center} PRESENTED AT\end{center}\bigskip \begin{center}\begin{large}}{\end{large}\end{center} \end{quotation}}
\def\Acknowledgements{\bigskip  \bigskip \begin{center} \begin{large} \bf ACKNOWLEDGEMENTS \end{large}\end{center}}

\def\beq{\begin{equation}}
\def\eeq#1{\label{#1}\end{equation}}
\def\eeqn{\end{equation}}

\def\beqa{\begin{eqnarray}}
\def\eeqa#1{\label{#1}\end{eqnarray}}
\def\eeqan{\end{eqnarray}}

\let\bar=\overbar

\def\Dslash{\not{\hbox{\kern-4pt $D$}}}
\def\dslash{\not{\hbox{\kern-2pt $\del$}}}

\def\msb{{\bar{\ssstyle M \kern -1pt S}}}

\usepackage{amsmath}
\newcommand*{\TeV}{\ensuremath{\text{Te\kern -0.1em V}}}
\newcommand*{\GeV}{\ensuremath{\text{Ge\kern -0.1em V}}}
\newcommand*{\pt}{\ensuremath{p_{\text{T}}}}
\usepackage{subfig}
\usepackage{placeins}
\usepackage{lineno}

\begin{document}
\begin{titlepage}
\pubblock

\vfill
\Title{Overview of searches for single production of vector-like top and bottom quarks with the ATLAS experiment at 13 $\TeV$}
\vfill
\Author{Johannes Erdmann \\ on behalf of the ATLAS Collaboration\footnote{Copyright 2018 CERN for the benefit of the ATLAS Collaboration. Reproduction of this article or parts of it is allowed as specified in the CC-BY-4.0 license.}}
\Address{\institute}
\vfill
\begin{Abstract}
Vector-like quarks (VLQ) that couple preferentially to third-generation Standard Model (SM) quarks are a well-motivated extension of the SM, which could solve in particular the hierarchy problem. While VLQs can be produced in pairs via the strong interaction at the LHC, the cross section for single production may be larger than the pair production cross section for VLQ masses above the current exclusion limits. An overview of the searches for the single production of VLQs at the ATLAS experiment that use data taken at a centre-of-mass energy of 13 TeV, corresponding to up to 79.8~fb$^{-1}$, is presented. Vector-like top quarks are searched for in the decay channel $T\rightarrow Wb$ and in the decay channel $T\rightarrow Zt$ with the $Z$ boson decaying to two charged leptons. Vector-like bottom quarks are searched for in the decay channel $B\rightarrow Hb$ with the Higgs boson decaying to two photons. No significant excess over the SM prediction is found and 95\% CL exclusion limits are set on the single production of VLQs as a function of their mass and coupling to SM quarks.
\end{Abstract}
\vfill
\begin{Presented}
$11^\mathrm{th}$ International Workshop on Top Quark Physics\\
Bad Neuenahr, Germany, September 16--21, 2018
\end{Presented}
\vfill
\end{titlepage}
\def\thefootnote{\fnsymbol{footnote}}
\setcounter{footnote}{0}

\section{Introduction}

Vector-like quarks (VLQ) are predicted in many extensions of the Standard Model (SM). Particular examples are models that predict a new strong interaction and solve the hierarchy problem by introducing the Higgs boson as the pseudo-Nambu--Goldstone boson of the explicitly broken symmetry of that interaction. While VLQ can be pair-produced at the LHC via the strong interaction, at large VLQ masses the cross sections for single-VLQ production may be larger than for pair production. However, the single-production cross section depends on the coupling of the VLQ to SM particles. This coupling $C$~\cite{Wulzer} can be expressed in terms of the branching ratios (BR) of VLQ and a generalized coupling $\kappa$~\cite{Buchkremer}, or in terms of the mixing angle with SM quarks~\cite{handbook}. In the searches discussed here, only mixing with third-generation SM quarks is assumed, allowing the decays of vector-like $T$ ($B$) quarks to $Wb$ ($Wt$), $Zt$ ($Zb$) or $Ht$ ($Hb$). The BRs are predicted in certain VLQ multiplet representations. In the ($T$) singlet, these are approximately 50:25:25, and in the ($B$~$Y$) doublet, they are approximately 50:50:0. Searches for single-$T$ and single-$B$ production with 13~\TeV\ data taken by the ATLAS experiment~\cite{ATLAS} at the LHC are presented.

\section{Search for \boldmath $B\rightarrow Hb$ \unboldmath}

A search for the $B$ quark is performed in the decay mode to $Hb$, using the $H\rightarrow\gamma\gamma$ decay in 79.8~fb$^{-1}$ of data~\cite{Hb}. In the event selection, a Higgs-like diphoton system is required, as well as at least one forward jet, characteristic of single-VLQ production, and at least one $b$-tagged jet. Novel $b$-tagging is used that applies a fixed cut on the $b$-tagging discriminant value up to a jet \pt\ of 250~\GeV, and above 250~\GeV, the cut on the discriminant value is loosened as a function of \pt, so that a constant $b$-tagging efficiency is maintained. No reconstructed electrons or muons are allowed in the events, and the invariant mass of the diphoton system and the highest-\pt\ $b$-tagged jet, $m_{\gamma\gamma b}$, is required to be larger than 300~\GeV. In order to further separate the signal from the background, $|y^*| = 0.5 \cdot |y^{\gamma\gamma} - y^b|$ is required to be smaller than 1.1, where $y^{\gamma\gamma}$ ($y^b$) is the rapidity of the diphoton system (the highest-\pt\ $b$-tagged jet).

\begin{figure}[htb]
\centering
\includegraphics[width=0.49\textwidth]{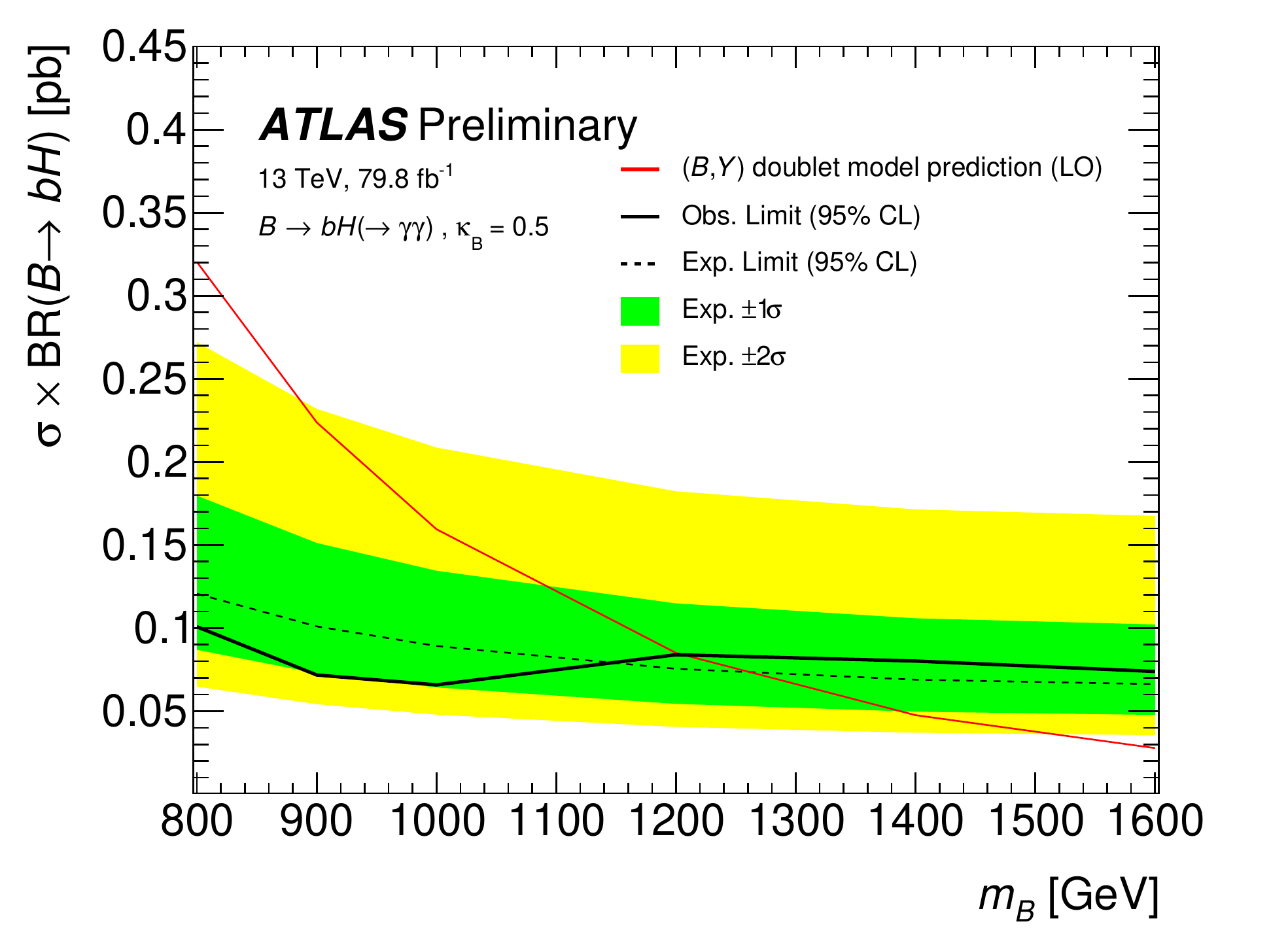}
\caption{Limits at 95\% CL on the cross section times BR to $Hb$ for the single production of a $B$ quark as a function of its mass, assuming a benchmark coupling value of $\kappa_B = 0.5$~\cite{Buchkremer}. Also shown is the expected cross section for the ($B$~$Y$) doublet model, which predicts a BR to $Hb$ of 50\%~\cite{Hb}.}
\label{fig:Hb}
\end{figure}

A resonance is searched for in the $m_{\gamma\gamma b}$ spectrum, where the main background process is continuum diphoton production, and a smaller contribution comes from events with a $H\rightarrow\gamma\gamma$ decay. The shape of the continuum background is estimated from data using events with an invariant diphoton mass in the range 105--120~\GeV\ or 130--160~\GeV. Its normalization is determined from the data in the search region. No significant excess above the SM background is found, and 95\% confidence level (CL) limits are set on the cross section times BR to $Hb$ as a function of the mass of the $B$ quark (Figure~\ref{fig:Hb}) for a benchmark coupling $\kappa_B = 0.5$~\cite{Buchkremer}. In the ($B$~$Y$) doublet model, $B$ quarks with masses smaller than 1210~\GeV\ are excluded.

\section{Search for \boldmath $T\rightarrow Wb$ \unboldmath}

A search for the $T$ quark is performed in the decay mode to $Wb$, using the $W\rightarrow\ell\nu$ decay in 3.2~fb$^{-1}$ of data~\cite{Wb}. In the event selection, exactly one electron or muon, missing transverse momentum larger than 120~\GeV, and a $b$-tagged jet with $\pt > 350~\GeV$ are required. Events are rejected if they contain additional high-\pt\ jets that are either close to the highest-\pt\ $b$-tagged jet or far away from it ($\Delta R < 1.2$ or $>2.7$). The distance in azimuth between the lepton and the highest-\pt\ $b$-tagged jet is required to be larger than 2.5 and at least one forward jet is required to be present.

\begin{figure}[htb]
\centering
\subfloat[]{\includegraphics[width=0.49\textwidth]{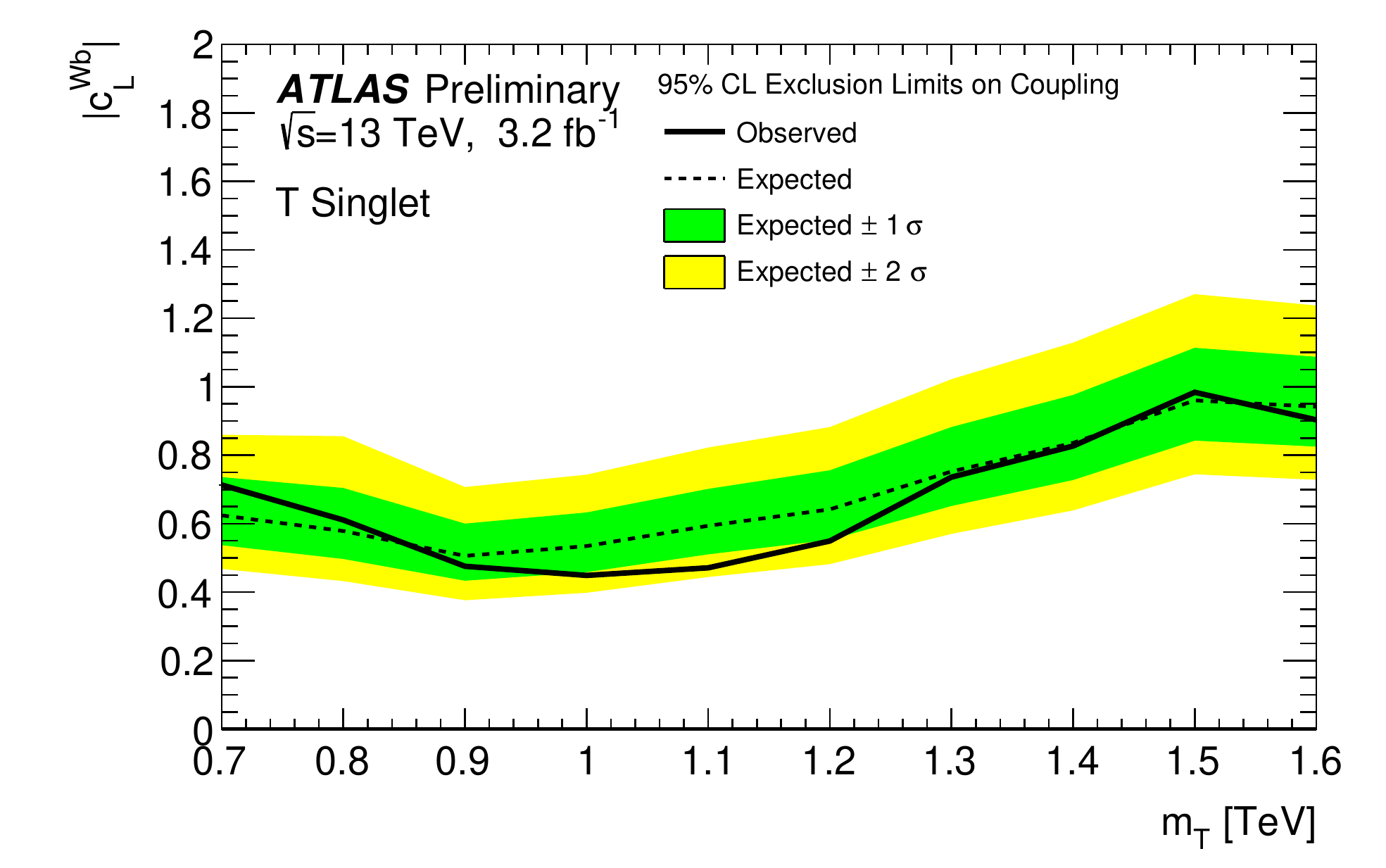}}
\subfloat[]{\includegraphics[width=0.49\textwidth]{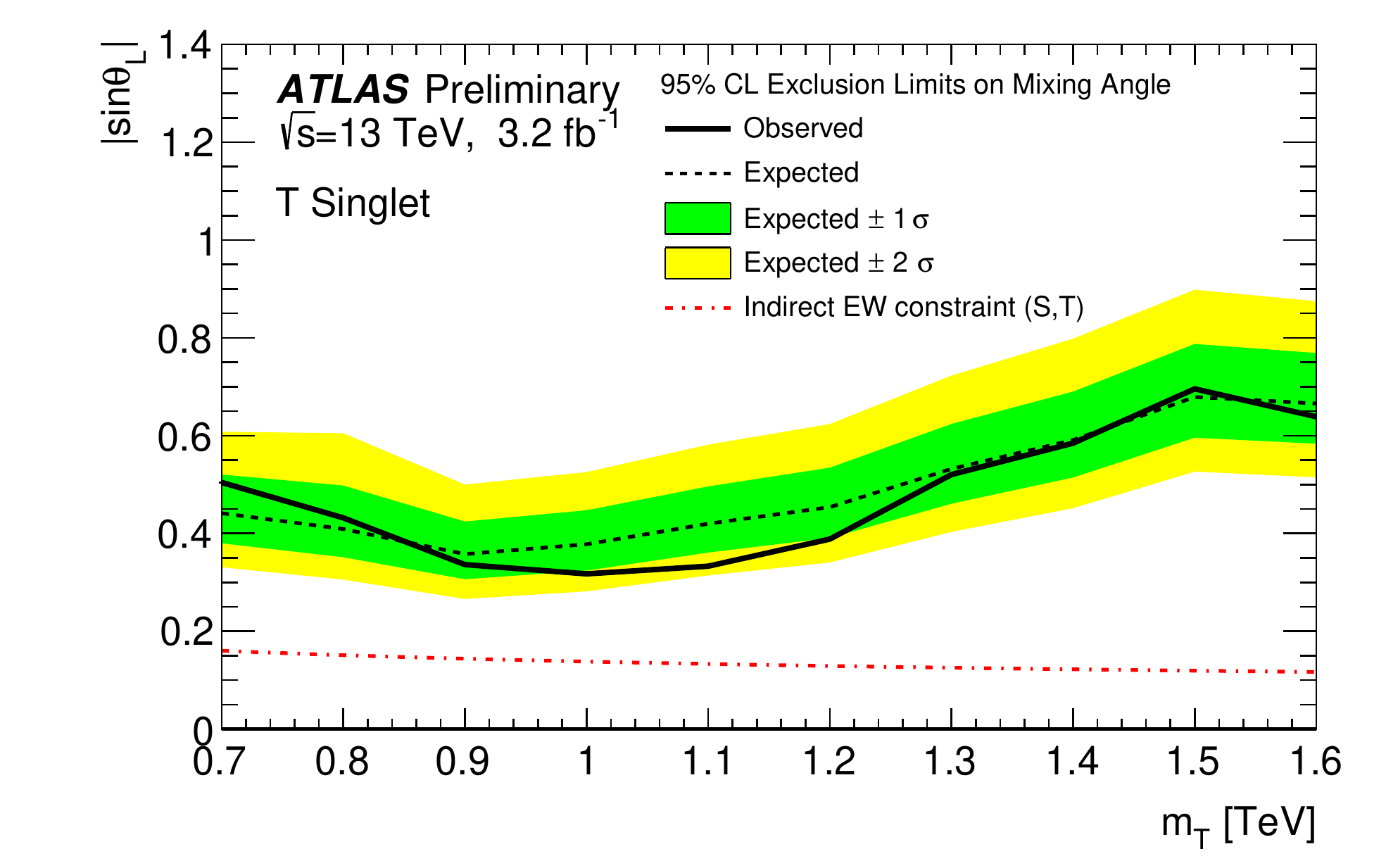}}
\caption{Limits at 95\% CL on (a) the left-handed coupling of the $T$ quark to the $W$ boson and the $b$-quark, and (b) the left-handed mixing angle of the $T$ quark and the top quark as a function of the mass of the $T$ quark in the singlet model~\cite{Wb}.}
\label{fig:Wb}
\end{figure}

The mass of the $T$ quark is reconstructed from the lepton, the missing transverse momentum, and the highest-\pt\ $b$-tagged jet, and a resonance is searched for in this spectrum, where the main background processes are $W$+jets, $t\bar{t}$ and single-top production. For $W$+jets and $t\bar{t}$ production, control regions (CR) are defined to validate and improve the modeling of these backgrounds. No significant excess above the SM background is found, and 95\% confidence level (CL) limits are set on the coupling and mixing angle of the $T$ quark as a function of its mass (Figure~\ref{fig:Wb}). As the coupling value does not only enter the production cross section but also the decay, the intrinsic VLQ width depends on the coupling value. When setting limits on the coupling or mixing angle, the change in width is taken into account by reweighting the MC samples, which were generated at a benchmark value of the coupling, using MC samples at truth level with a variable coupling value.

\section{Search for \boldmath $T\rightarrow Zt$ \unboldmath}

A search for the $T$ quark is performed in the decay mode to $Zt$, using the $Z\rightarrow\ell\ell$ decay in 36.1~fb$^{-1}$ of data~\cite{Zt}. Events are selected with two electrons or muons of opposite electric charge and with a mass close to the mass of the $Z$ boson ($Z$-boson candidate). Two channels are defined, which use events with exactly two charged leptons and events with at least three charged leptons, respectively. In both channels, at least one $b$-tagged jet is required in the event selection, as well as a forward jet. The \pt\ of the $Z$-boson candidate is required to be larger than 200~\GeV\ (150~\GeV) in the dilepton (trilepton) channel. In the dilepton channel, at least one top-tagged jet with a large radius parameter is required. In the trilepton channel, the \pt\ of the highest-\pt\ electron or muon that is not part of the $Z$-boson candidate must be larger than 200~\GeV. In both channels, a requirement is made in the event selection that suppresses the contribution from VLQ pair production, so that the focus of the analysis is solely on the single-production process.

\begin{figure}[htb]
\centering
\includegraphics[width=0.49\textwidth]{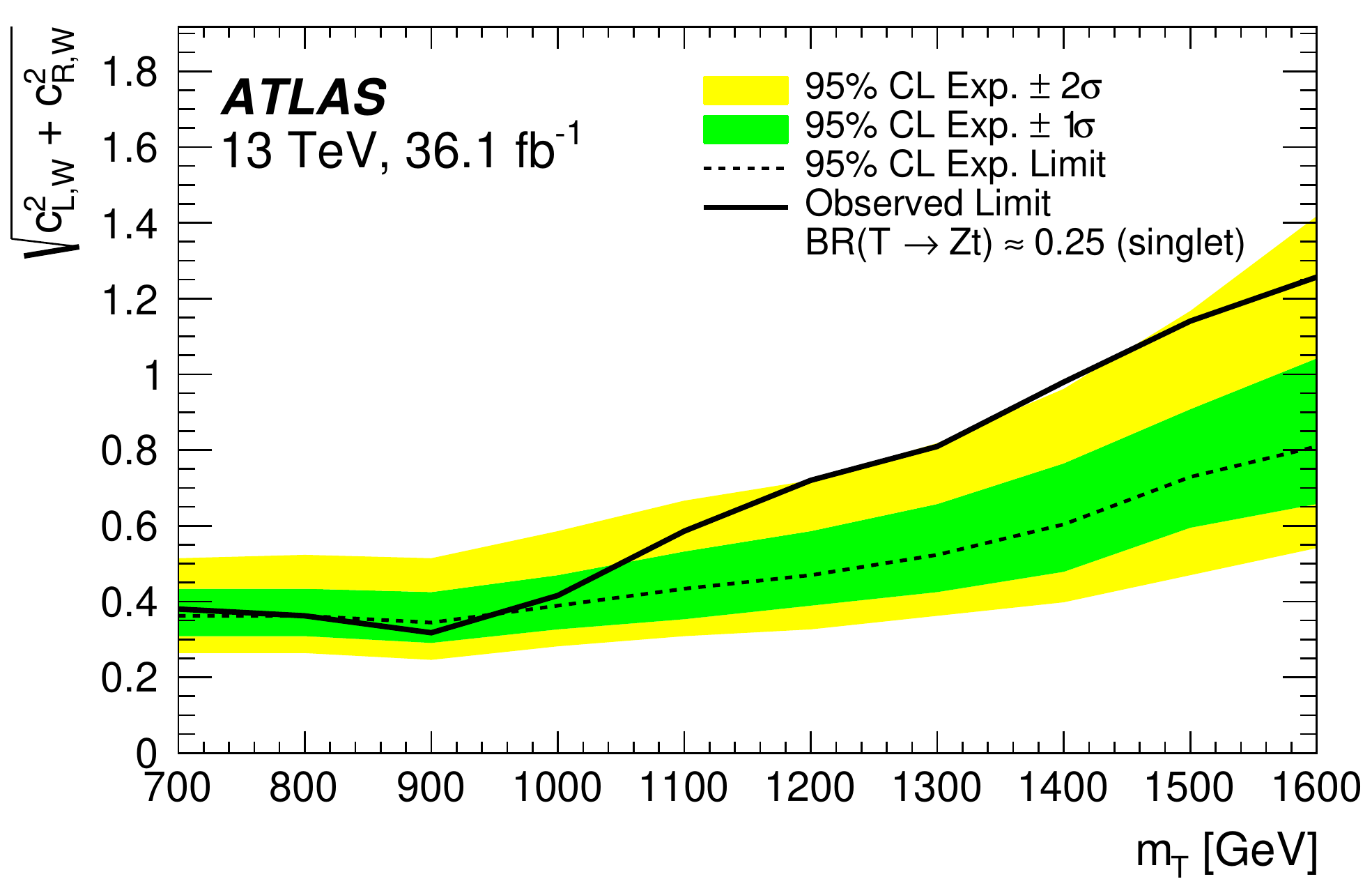}
\includegraphics[width=0.49\textwidth]{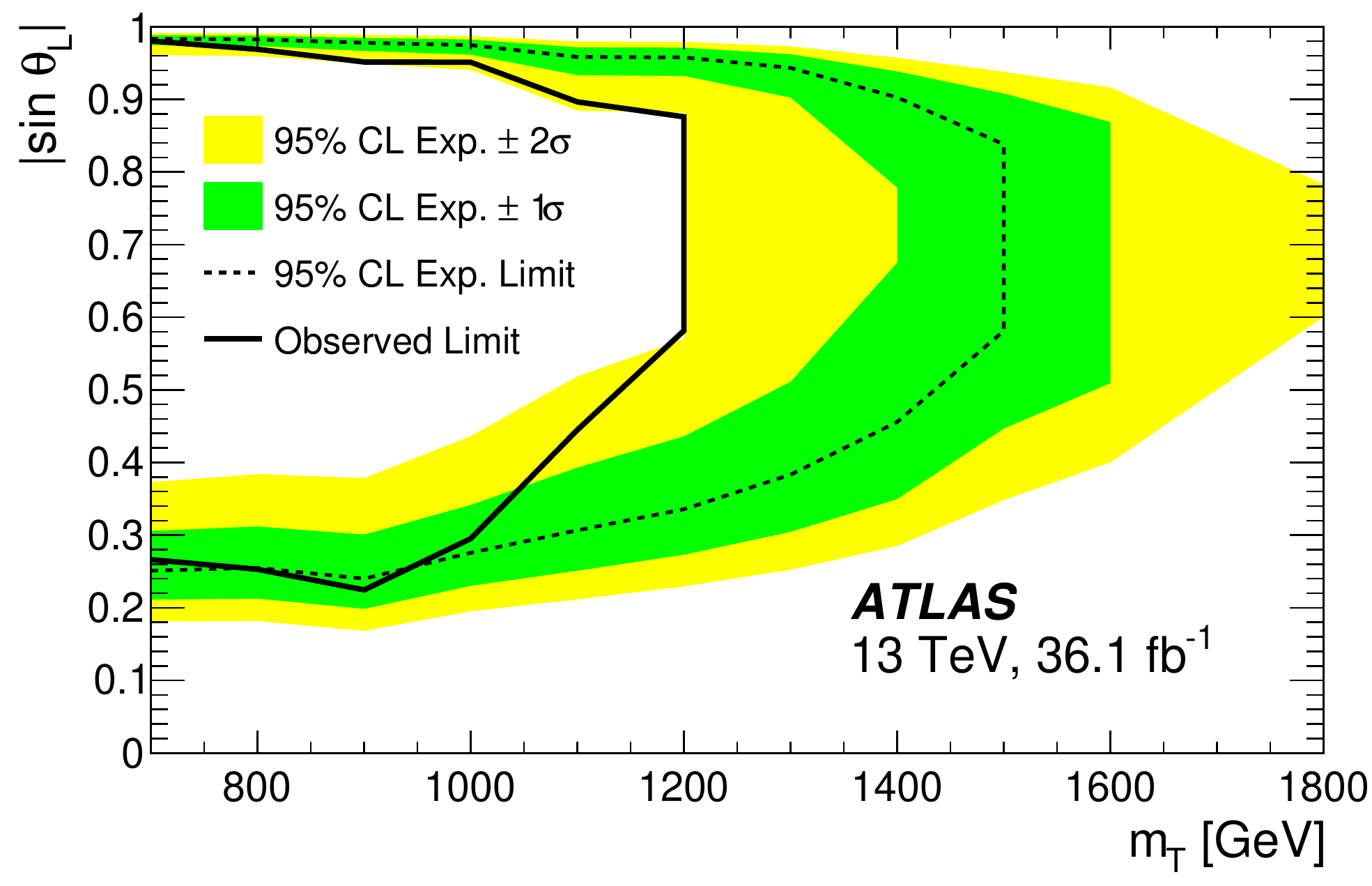}
\caption{Limits at 95\% CL on (a) the coupling of the $T$ quark to the $W$ boson and the $b$-quark, and (b) the left-handed mixing angle of $T$ quark and the top quark as a function of the mass of the $T$ quark in the singlet model~\cite{Zt}.}
\label{fig:Zt}
\end{figure}

A resonance is searched for in the spectrum of the invariant mass of the highest-\pt\ top-tagged large-radius jet and the $Z$-boson candidate (the scalar sum of the transverse momenta of all jets and electrons and muons) in the dilepton (trilepton) channel. The main background processes are $Z$+jets (diboson and $t\bar{t}X$, where $X$ is a $Z$ boson in a large fraction of the events) production in the dilepton (trilepton) channel. For these processes, CRs are defined to validate and improve the modeling of the backgrounds. No significant excess above the SM background is found, and 95\% confidence level (CL) limits are set on the coupling and mixing angle of the $T$ quark as a function of its mass (Figure~\ref{fig:Zt}). As in the $T\rightarrow Wb$ analysis, the MC samples were reweighted to set limits as a function of the coupling or mixing angle.

\FloatBarrier

\section{Conclusions}
Single production of vector-like quarks was searched for in proton--proton collision data taken at the LHC with the ATLAS detector at a center-of-mass energy of 13~\TeV, using data corresponding to integrated luminosities of 3.2--79.8~fb$^{-1}$. No significant excess above the Standard Model background predictions was found, and 95\% confidence level limits were set on the mass of the $B$ quark, and on the coupling and mixing angles of $T$ quarks in benchmark models.

\Acknowledgements
The author would like to acknowledge the support by BMBF, Germany (FSP-103).

\end{document}